\documentclass[%
 reprint,
 amsmath,amssymb,
 aps,
prd,
showkeys
]{revtex4-1}

\usepackage{dcolumn}
\usepackage{diagbox}
\usepackage{graphicx} 
\usepackage{amsmath}
\usepackage{amssymb}
\usepackage{physics}
\usepackage{color}
\usepackage{float}
\usepackage{yhmath}
\usepackage[dvipsnames]{xcolor}
\usepackage[normalem]{ulem}
\usepackage{listings}

\usepackage{hyperref}

\allowdisplaybreaks

\begin{document}

\title{A Gravitational Waveform Model for Detecting Accelerating Inspiraling Binaries}
\author{Malcolm Lazarow}
\email{mlazarow@berkeley.edu}
\affiliation{Department of Physics, University of California, 366 Physics North MC 7300, Berkeley, CA. 94720, USA}
\author{Nathaniel Leslie}
\affiliation{Department of Physics, University of California, 366 Physics North MC 7300, Berkeley, CA. 94720, USA}
\author{Liang Dai}
\affiliation{Department of Physics, University of California, 366 Physics North MC 7300, Berkeley, CA. 94720, USA}
\date{\today}

\begin{abstract}
    We present an analytic frequency-domain gravitational waveform model for an inspiraling binary whose center-of-mass undergoes a small acceleration, assumed to be constant during the detection, such as when it orbits a distant tertiary mass.
    The center-of-mass acceleration along the line of sight is incorporated as a new parameter that perturbs the standard TaylorF2 model. We calculate the wave phase to 3\textsuperscript{rd} post-Newtonian order and first order in the acceleration, including the effects of aligned component spins. It is shown that acceleration most significantly modifies the wave phase in the low frequency portion of the signal, so ground-based detectors with a good sensitivity at low frequencies are the most effective at detecting this effect. We present a Fisher information calculation to quantify detectability at Advanced LIGO A+, Cosmic Explorer, and Einstein Telescope over the mass range of neutron stars and stellar-mass black holes, and discuss degeneracy between acceleration and other parameters. We also determine the parameter space where the acceleration is large enough that the wave phase model would have to be extended to nonlinear orders in the acceleration.
\end{abstract}
\maketitle
\section{Introduction}
\renewcommand{\arraystretch}{1.1}
\setlength{\tabcolsep}{5pt}

Gravitational waves (GWs) from inspiraling and merging compact binaries have been detected in large quantities by the LIGO-Virgo-KAGRA (LVK) collaboration \cite{gwosc_list}. This experimental success is enabled not only by improvement of detector sensitivity \cite{Creighton_2003, Martynov_2016, Sathyaprakash_2009}, but also by advances in developing highly accurate analytic \cite{Damour_2014, Blanchet_2014} and numerical \cite{Centrella_2010, Sperhake_2015} gravitational waveform models for the two-body problem in general relativity (GR).

The common approach for GW detection and parameter inference \cite{Christensen_2022, Jaranowski_2012, Sathyaprakash_2009} requires readily available waveform models that follow the binary's motion to high accuracy over many orbits \cite{Owen_1999, Owen_1998, Roulet_2019}. Progress in the GR two-body problem has enabled rapid generation of accurate waveform models for bound two-body orbits, based on which robust, scalable analyses of the strain time series at interferometers have been implemented \cite{Owen_1999, Berti_2007, London_2014}. On the other hand, discovery of novel GW sources can be hindered by a lack of accurate waveform models.

One type of system that is still absent from the current GW source catalog are hierarchical triple systems, where two compact objects in a bound orbit themselves orbit a third distant mass \cite{Eggleton_1995, Toonen_2016, Naoz_2013, Randall_2018}.
Accurately calculating secularly evolving three-body orbits in GR and rapidly generating the corresponding waveform templates for the entire three-body system is a topic of ongoing numerical investigations ~\cite{Bonetti_2017, Galaviz_2011}. The inspiralling two-body GW signal, however, can be treated perturbatively as its motion secularly evolves due to the third body.

The focus of this paper will be gravitational waves from a simple case of hierarchical triple systems. Our investigation is motivated by the detection of GW190814 \cite{GW190814}, an inspiral-merger event of a $23\,M_{\odot}$ black hole (BH) and a $2.5\,M_{\odot}$ compact object. The nature of the less massive compact object is puzzling: it is either the lightest BH known to exist or an exceptionally massive neutron star (NS). To offer an explanation for how GW190814 might have formed, some recent studies (e.g. \cite{Hamers_2019, Lu_2020}) have investigated if it might be the end state of a hierarchical triple system, in which two NSs merged to form the $2.5\,M_{\odot}$ object.  This setting may also be relevant for events with similar mass ratios with a component in the mass gap (e.g. GW200210 ~\cite{o3b}).  In Ref.~\cite{Lu_2020}, it is argued that roughly 10\% of binary NS mergers might occur near a massive third body if such a triple mechanism is the origin of mergers that involve a mass-gap object.  If the signature of a tertiary mass around merging NS binaries can be detected in the GW signal, it would be a discovery of three-body interactions as a dynamical origin of these LVK GW sources.

There is also motivation to study binary BHs (BBHs) whose center-of-mass (CoM) accelerates during inspiral events. Studies of BBH formation in active galactic nuclei and globular clusters often predict nontrivial CoM motions~\cite{Stone_2016, McKernan_2018, Bartos_2017, Antonini_2016, Yang_2019, Tiwari_2023}. Studies of single-binary scattering have demonstrated that GW sources can form during chaotic three-body interactions~\cite{Samsing_2014, Samsing_2018a, Ginat_2022}.  The hypothesis that events like GW190814 arise from hierarchical triples hinges on double merger events, which have also been observed in single-binary BH scattering simulations~\cite{Samsing_2018b}.

In this paper, we present an analytic gravitational waveform model for two inspiraling compact objects whose CoM has a small acceleration due to the gravitational pull of a tertiary body. CoM acceleration imprints a non-trivial distortion to the wave phase, whereas a constant CoM velocity simply imparts a Doppler shift and is fully degenerate with the orbital frequency, source distance, and source masses~\cite{Cutler_1994}. We specialize to the case in which the third body is sufficiently distant compared to the compactness of the binary so that the inspiraling binary as a whole simply follows two-body geodesic motion. In the free-falling frame of the binary, the internal inspiral motion of the binary is assumed to be well described by the usual two-body dynamics in GR. In this situation, tertiary tidal perturbations on the binary orbital phase is assumed to be negligible over LVK signal timescales, and hence the CoM acceleration manifests as a time-dependent delay for the detected GWs.

There are previous studies that address CoM acceleration for GW sources, but our full analytic waveform model has yet to appear in the literature. Accelerating sources have been considered in sources for the Laser Interferometer Space Antenna (LISA); for example, concerning extreme-mass-ratio-inspirals near a supermassive BH ~\cite{Yunes_2011}, comparable mass binaries embedded in an environment ~\cite{Iyanoshi_2017, Randall_2019, Wong_2019}, double white-dwarf binaries near a tertiary star ~\cite{Xuan_2021}, a compact binary wobbling due to a circumbinary planet~\cite{Tamanini_2019}, and apparent acceleration of a source due to a time-dependent redshift ~\cite{Tamanini_2020}. 
 The Newtonian model presented in ~\cite{Tamanini_2019} has been used in a recent parameter estimation study for GW190814~\cite{Han_2024}, though we demonstrate that higher order post-Newtonian (PN) corrections are needed for a complete analysis for ground-based and next-generation detectors. For ground-based compact object binaries, Ref.~\cite{Meiron_2017} gives a back-of-envelope estimation for the detectability of a tertiary's imprint from the GW signal within a hierarchical triple. 
 Another recent study has outlined a numerical framework for analyzing time-dependent Doppler effects for full inspiral-merger-ringdown models~\cite{Chamberlain_2019}. We instead consider detection at ground-based GW detectors for inspiral events, which requires higher-order post-Newtonian (PN) corrections to the orbital phase.  Our model reduces to the Newtonian result~\cite{Tamanini_2020} in the appropriate limit. 

While we were independently working on this manuscript, another study
carried out an in-depth investigation on the detectability of CoM acceleration at current and upcoming ground-based GW detectors~\cite{Vijaykumar_2023}. The authors provided 3.5PN analytic waveforms accurate to first order in CoM acceleration for non-spinning compact objects, and presented Fisher forecasts and Bayesian parameter inference on GW170817 and GW190425.
Their analytic results agree with ours, although we will present new analytic waveforms that include the effects from component spins aligned with the orbital angular momentum. In addition, we will provide analytic formulae for the Cram\'er-Rao (CR) bound, quantify the number of observed orbital cycles due to CoM acceleration for all given PN orders, and demonstrate that the CoM acceleration is degenerate with the component spins at a comparable level with the chirp mass and the reduced mass. We explicitly demonstrate that neglecting spin would decrease the CR bound on CoM acceleration by a factor of two. Doing so could lead to false detection, even if the inferred spin is centered around zero. In summary, we rigorously quantify, using Fisher information, the detectability of the CoM acceleration effect at forthcoming ground-based GW observatories (Advanced LIGO A+, Cosmic Explorer, and Einstein Telescope) and present analytical formulae that are broadly applicable and have yet to appear in the literature.


The paper is organized as follows. In Section \ref{sec:model}, we present an analytic, frequency-domain model for the GW phase, which incorporates the CoM acceleration of the inner binary induced by a distant tertiary up to the third PN order. In Section \ref{sec:statistical_background}, we present a statistical framework based on Fisher information in which we forecast the detectability of CoM acceleration. The results for forthcoming ground-based GW detectors will be given in Section \ref{sec:detectability_analysis}. We will give concluding remarks in Section \ref{sec:concl}. We set $G=c=1$ throughout.

\section{Model Incorporating Small Center-of-Mass Acceleration}
\label{sec:model}

This section presents our GW phase model, which incorporates the CoM acceleration of an inspiraling binary on a circular orbit, embedded in a hierarchical triple system.
The relative motion of the two bodies, which decouples from the CoM acceleration because the binary CoM is in free fall, is given by the standard two-body GR dynamics. We use the analytic PN phase of the standard two-body problem \cite{Blanchet_2014} to derive the observed wave phase.


The inspiraling binary has a line-of-sight acceleration:
\begin{equation}
    a_{\parallel} = (M_3/r^2)\,\hat{l}\cdot\hat{r},
\end{equation}
where $M_3$ is the tertiary mass, $r$ is the distance between the binary and the tertiary, $\hat{l}$ is a unit vector along the line of sight, and $\hat{r}$ is a unit vector pointing from the binary CoM to the tertiary. The GW signals from inspiraling compact binaries detected by ground-based observatories would last for no more than $\mathcal{O}(10^3)$ seconds, so we assume that $a_{\parallel}$ is approximately constant during the event.

We now correct the observed time-domain orbital phase due to $a_\parallel$:
\begin{subequations}
\begin{align}
    \label{eq:orbital_phase_model}
    \phi_{\alpha}(t) &= \phi_0\left(t + \alpha (t-T)^2\right),\\
    \label{eq:alpha_value}
    \alpha &= \frac{a_\parallel}{2} = 9.9\times 10^{-12}\,\text{s}^{-1}\, \left(\frac{M_3}{M_{\odot}}\right)\left(\frac{1\text{ AU}}{r}\right)^2,
\end{align}
\end{subequations}
where $\phi_\alpha(t)$ and $\phi_0(t)$ are the orbital phases with and without the CoM acceleration, respectively, and we have restored physical units in $\alpha = a_{\parallel}/2c$ for the second line. The time constant $T$ corresponds to the choice of a reference time. See Appendix \ref{sec:derivation} for details on how $T$ is chosen and how $\phi_{\alpha}(t)$ can be evaluated as a perturbative series in $\alpha$.

The frequency-domain waveform is the Fourier transform of the time-domain GW waveform~\cite{Cutler_1994},
\begin{equation}
    \label{eq:t_waveform}
    h_{\alpha}(t) = {\rm Re}\left[ A(t)\, e^{2\,i\,\phi_{\alpha}(t)} \right].
\end{equation}
In this work, we neglect corrections to the wave amplitude $A(t)$ due to CoM acceleration, although it has been calculated to first order in $\alpha$~\cite{Tamanini_2019}. See Ref.~\cite{Buonanno_2009} and references therein for details on how the time dependence of the amplitude can be treated.

We analytically calculate the Fourier transform using the stationary phase approximation (SPA). This approach results in the widely used ``Taylor" models, in which the orbital phase of the binary is obtained by integrating the bound state energy divided by the gravitational radiation flux, written as a PN expansion, and then using the SPA (see \cite{Blanchet_2014, Buonanno_2009, Boyle_2008, Poisson_2014} and Appendix \ref{sec:derivation} for further details). The result is
\begin{multline}
\label{eq:model}
    \psi_{\alpha}(f) = 2\pi f \,t_c - \phi_c - \frac{\pi}{4}\\
    +x^{-5/2}\, \sum_{n=0}^6 \underbrace{x^{n/2}\,\left(P_{n} + S_n + \alpha\,A_n\, x^{-4}\right)}_{n/2\text{ PN}}\\
    + \mathcal{O}(\underbrace{\alpha^2}_{\text{``quadratic",}}, \underbrace{x}_{\geq\text{3.5 PN}}),
\end{multline}
where $x = \left(\pi\,M\,f\right)^{2/3}$ is the PN expansion parameter, the underbrace in the sum denotes the PN order of the term, and the ``quadratic" underbrace below $O\left(\alpha^2\right)$ denotes higher order corrections in $\alpha$. The time constant $t_c$ and phase constant $\phi_c$ correspond to the point of coalescence. 

The coefficients $P_{n}, S_{n},$ and $A_{n}$ are given explicitly in Appendix \ref{sec:derivation} (see Eq.~\eqref{eq:coefficient}). In the expressions for these coefficients are the binary's (redshifted) reduced mass $\mu = \left(1 + z\right)\,m_1\,m_2/M$, the (redshifted) chirp mass $\mathcal{M} = \left(1+z\right)\,\mu^{3/5}\,M^{2/5}$, and $-1\leq\chi_1, \chi_2 \leq 1$ are dimensionless spins of the binary, each assumed to be parallel to the orbital angular momentum for simplicity. Here $M=m_1+m_2$ is the binary total mass.

The frequency-domain GW waveform is then
\begin{equation}
\label{eq:total_waveform}
    h(f) = \mathcal{A}\,f^{-7/6}\,e^{i\,\psi_{\alpha}(f)},
\end{equation}
where $\mathcal{A}$ is an overall amplitude normalization constant.
The factor of $f^{-7/6}$ comes from the SPA and dependence on the chirp mass has been absorbed into $\mathcal{A}$ (see Appendix \ref{sec:derivation} as well as \cite{Cutler_1994, Buonanno_2009} and references therein). From Eq. \eqref{eq:model}, we see that $\alpha$ affects the entire frequency spectrum of the GW phase, whereas sky position manifests as a constant phase change, which is inferred by comparing strain signals recorded in multiple detectors~\cite{Cutler_1994}. For simplicity, in the following discussion we treat the GW detection process as if there is only one detector. To account for multiple detectors, we can rescale the amplitude constant $\mathcal{A}$ to match the network SNR. In this way, we essentially assume that the CoM acceleration does not affect the inference of extrinsic parameters such as sky location and binary position angle.

We now discuss three features and caveats of this analytic waveform model. First, the corrections due to acceleration appear at $\mathcal{O}\left(f^{-8/3}\right)$ relative to the non-acceleration terms. The effect of CoM acceleration on the phase is therefore most pronounced at low frequencies. The importance of the low frequency GW signal can be understood by calculating the number of wave cycles that arise from contributions of the usual TaylorF2 terms, $P_{n}$ and $S_{n}$, and from acceleration, $A_{n}$, as a function of the minimum frequency observed at each order in the PN expansion (see Eq. \eqref{eq:phi_t} for how one derives the observed orbital phase $\phi_{\alpha}(f)$).

The number of cycles is defined by
\begin{align}
\label{eq:n_cycles}
    \mathcal{N} &= \frac{\phi_{\alpha}\left(f_{\rm ISCO}\right) - \phi_{\alpha}\left(f_{\rm min}\right)}{\pi},
\end{align}
where $f_{\rm ISCO} = 1/6^{3/2} \pi M$ is the orbital frequency at an approximate innermost stable circular orbit (ISCO). The values for $\mathcal{N}$ are given in Table \ref{tab:num_cycles}. There, we calculate the number of cycles, at each order in the PN expansion, for two values of $f_{\rm min}$, 10 Hz and 1 Hz. Each column lists the contributions from the $P_{n}$, $S_{n}$, and $A_{n}$ terms. 

For $f_{\rm min} = 10\,$Hz, the frequency at which seismic noise is expected to limit the sensitivity of advanced LIGO \cite{Aasi_2015}, acceleration contributes less than one cycle unless $\alpha \gtrsim 10^{-7}\,\text{s}^{-1}$; however, our model, accurate to first order in $\alpha$, would not be valid for such large $\alpha$ values for binary NS inspirals (see Appendix \ref{sec:linear_alpha_justification} and Figure \ref{fig:results_all_params}). If the lowest frequency is set to $1\,$Hz, acceleration then contributes tens of thousands of detectable orbital cycles for $\alpha \leq 10^{-7}\,{\rm s}^{-1}$ and many cycles for higher orders in the PN expansion, so they are necessary for a complete parameter inference study. The more sensitive a GW detector is at low frequencies, the more capable it is to detect CoM acceleration, which implies that third-generation detectors are much more promising than current LVK detectors for CoM inference. Our results for $f_{\rm min} = 10\,$Hz reproduce Tables 3 and 4 in \cite{Blanchet_2014} and Table 1 in \cite{Poisson_1995}. We further note that decreasing $f_{\rm min}$ from $10\,$Hz to $1\,$Hz results in at least an order of magnitude more observable cycles from the $P_n$ and $S_n$ terms, but four to five orders of magnitude increase in the CoM acceleration terms. The spin dependent terms in $A_n$ also result in many cycles, so they must be included to accurately model the waveform~\cite{Cutler_1993}.

\begin{table*}
    \centering
    \begin{tabular}{|c||c|c|c|}
        \hline
        PN Order & $P_{n}$ & $S_{n}$ & $A_{n}$\\
        \hline\hline
        & \multicolumn{3}{|c|}{$f_{\rm min} = 10\text{ Hz}$}\\
        \hline
        0 & $1.6\times 10^{4}$ & - & $5.4\times 10^{7}\alpha$\\
        1 & $4.4\times 10^2$ & - & $2.1\times 10^6\alpha$\\
        1.5 & $-2.1\times 10^2$ & $66\left(\chi_1 + \chi_2\right)$ & $\left(-7.5 + 2.4\,\chi_1 + 2.4\,\chi_2\right)\times 10^5\alpha$\\
        2 & $9.9$ & - & $4.4\times 10^4\alpha$\\
        2.5 & $-11$ & $9.3\left(\chi_1 + \chi_2\right)$ & $\left(-3.1 + 1.8\,\chi_1 +1.8\,\chi_2\right)\times 10^4\alpha$\\
        3 & $2.6$ & $-4.0\left(\chi_1 + \chi_2\right)$ & $\left(4.4 - 4.6\,\chi_1 - 4.6\,\chi_2\right)\times 10^3\alpha$\\
        \hline\hline
        & \multicolumn{3}{|c|}{$f_{\rm min} = 1\text{ Hz}$}\\
        \hline
        0 & $7.4\times 10^5$ & - &$ 1.2\times 10^{12}\alpha$\\
        1 & $4.4\times 10^3$ & - & $9.6\times 10^9\alpha$\\
        1.5 & $-1.0\times 10^3$ & $3.1\times 10^2\left(\chi_1 + \chi_2\right)$ & $\left(-16 + 5.1\,\chi_1 + 5.1\,\chi_2\right)\times 10^8\alpha$\\
        2 & $24$ & - & $4.4\times 10^7\alpha$ \\
        2.5 & $-17$ & $13\left(\chi_1 + \chi_2\right)$ & $\left(-14 + 8.3\,\chi_1 +8.3\,\chi_2\right)\times 10^6\alpha$\\
        3 & $2.7$ & $-4.5\left(\chi_1 + 4.5\,\chi_2\right)$ & $\left(0.81 - \chi_1 -\chi_2\right)\times 10^6\alpha$\\
        \hline
    \end{tabular}
    \caption{Contribution to the number of observed cycles as a function of $\alpha$ [s$^{-1}$], calculated via Eqs. \eqref{eq:n_cycles} and \eqref{eq:phi_t}, of the orbital phase for a binary with $m_1 = m_2 = 1.4 M_{\odot}$ for two values of $f_{\rm min}$.  The left-most column denotes the PN order of the contributing waveform cycles, while the second, third, and fourth columns denote the contribution from the $P_{n}$, $S_{n}$, and $A_{n}$ terms, respectively, see Eqs. \eqref{eq:model} and \eqref{eq:coefficient} (note that there are both spin-independent and dependent terms within $A_{n}$).  We consider two values for $f_{\rm min}$, 10 Hz in the first tabular block and 1 Hz in the second.  The highest frequency for the orbit is taken to be $f_{\rm ISCO} = 1/\left(6^{3/2}\pi M\right)$, the ISCO for a massive test particle in Schwarzschild spacetime, for all cases.  It is clear that the number of cycles arising from CoM acceleration dramatically increases as the minimum frequency is decreased. Therefore, sensitivity improvements at low frequencies greatly enhance the detectability of $\alpha$.  For the lowest minimum frequency, $f_{\rm min} = 1\text{ Hz}$, CoM acceleration can contribute at least one cycle so long as $\alpha \gtrsim 10^{-12}\text{ s}^{-1}$.  We see also that the 2.5 PN terms can contribute a cycle for $\alpha\sim 10^{-7}\text{ s}^{-1}$, but accelerations that are much larger requires considering $O(\alpha^2)$ in Eq. \eqref{eq:model}, see Appendix \ref{sec:linear_alpha_justification}. 
 Note these values are independent of detector sensitivity.}
    \label{tab:num_cycles}
\end{table*}

Second, we have assumed binary inspiral during the entire duration of the GW signal. Since the merger and ringdown phases are very shortlived compared to inspiral, the latter phase provides almost all the information about CoM acceleration. For BBHs, it has been shown that using inspiral models for full inspiral-merger-ringdown (IMR) events yields parameter posteriors that are consistent with using IMR models \cite{Ghosh_2016, Ghosh_2017}.  Other studies have demonstrated how one would carry out a numerical analysis of a time-dependent phase delay, similar to ours, for full IMR events~\cite{Chamberlain_2019}. Our simplifying treatment should be valid except for the very massive BBH events.

Third, we have only incorporated the effect of acceleration at the linear order in $\alpha$. Our model, Eq.~\eqref{eq:model}, is of the form $\psi_{\alpha}(f) = \psi_0(f) + \alpha\,\psi_1(f) + \alpha^2\,\psi_2(f) + \dots$, with $\psi_2(f)$ and all higher-order pieces neglected. This is valid provided $|\alpha| \ll |\psi_1(f)/\psi_2(f)|$. We investigate this assumption in Appendix \ref{sec:linear_alpha_justification}. We find that for binaries with $m_1 = m_2 = 1.4\,M_{\odot}$, the quadratic term becomes important only for rather large acceleration values $\alpha \gtrsim 6\times 10^{-7}\,{\rm s}^{-1}$.

We are now in a position to address the data analysis of ~\cite{Han_2024}, in which a CoM acceleration posterior with mean $\alpha \approx 10^{-3}\text{ s}^{-1}$ and standard deviation $\sim 10^{-3}\text{ s}^{-1}$ are inferred from GW190814.  This result is consistent with zero CoM acceleration and therefore is compatible with the null hypothesis that the source is not undergoing CoM acceleration.  The signal entered the LIGO Livingston detector at around $30$ Hz~\cite{GW190814}.  According to Eq. \eqref{eq:n_cycles}, the number of cycles contributed by CoM acceleration for an event with $f_{\rm min} = 30\,\text{ Hz}$ is $2.13\times 10^{2}\,\alpha$ at the Newtonian order and $7.95\times 10\,\alpha$ at 1 PN order.  This is an insufficient number of cycles to detect astrophysically plausible values $\alpha \ll 10^{-3}\,{\rm s}^{-1}$, consistent with the result of ~\cite{Han_2024}.

In the remainder of the paper, we forecast the precision with which $\alpha$ can be measured from GW data. To this end, we discuss the Fisher information formalism and detector sensitivities in the next Section.

\section{Framework for Parameter Estimation}
\label{sec:statistical_background}

We now summarize the framework we use to predict the precision of parameter inference using our model.  In Section \ref{sec:Fisher_framework}, we introduce the Fisher information matrix.  This enables enables us to forecast a lower bound for posterior variances that would be obtained by Markov-Chain Monte-Carlo (MCMC) techniques \cite{Christensen_2022}. Detector sensitivities for the third-generation detectors of interest are given in Section \ref{sec:PSDs}.

\subsection{Fisher Information and the Posterior Covariance}
\label{sec:Fisher_framework}

The Fisher information formalism has been widely used to quantify parameter inference precision in GW science. We refer the readers to \cite{Finn_1992, Poisson_1995, Cutler_1994, Jaranowski_2012, Sathyaprakash_2009} and references therein for further information and we proceed by defining quantities needed for our analysis.

When a signal is present, GW strain data can be decomposed as, $s(f) = h\left(f;\hat{\theta}\right) + n(f)$, where $h\left(f; \hat{\theta}\right)$ is a GW waveform, $\theta$ is a placeholder for the entire set of source parameters, and $\hat{\theta}$ denotes the parameter values that best fit the data for given detector noise $n(f)$. The detector noise $n(f)$ is assumed to be stationary, colored Gaussian noise characterized by the one-sided power spectral density (PSD) $S_n(f)$. Because $\hat{\theta}$ depends on the noise realization $n(f)$, it is also a random variable with mean equal to the source's true parameter value \cite{Jaranowski_2012}.

For a sufficiently high signal-to-noise ratios (SNR), the likelihood function for the parameters $\theta^a$ has a Gaussian form
\begin{subequations}
\begin{align}
\label{eq:gaussianprob}
    L\left(\theta\big\rvert s\right) &\propto \exp\left[-\frac{1}{2}\,\Gamma_{ab}\,\delta\theta^a\,\delta\theta^b\right],\\
    \label{eq:Fisher}
    \Gamma_{ab} &= \left\langle \frac{\partial h}{\partial \theta^a}, \frac{\partial h}{\partial \theta^b}\right\rangle\bigg\rvert_{\theta = \hat{\theta}},\\
    \label{eq:inner_product}
    \left\langle \frac{\partial h}{\partial \theta^a}, \frac{\partial h}{\partial \theta^b}\right\rangle &= \Re\left[4\,\int_{0}^{\infty}\frac{\frac{\partial h}{\partial \theta^a}\frac{\partial h^*}{\partial \theta^b}}{S_n(f)}\,\dd f\right],
\end{align}
\end{subequations}

\noindent where $\Gamma_{ab}$ is the Fisher information matrix, $\delta\theta^c$ is the difference between a given parameter $\theta^c$ and the posterior's mean, $\langle\cdots\rangle$ denotes the standard inner product between two frequency domain signals, and $S_n(f)$ is the PSD of the detector under consideration \cite{Cutler_1994, Finn_1992, Poisson_1995, Sathyaprakash_2009, Christensen_2022}.  See \cite{Finn_1992, Cutler_1994} and references therein for details about this framework.

The likelihood function in Eq. \eqref{eq:gaussianprob} is a multivariate Gaussian whose covariance matrix is the inverse of the Fisher matrix,
\begin{equation}
    \label{eq:covariance}
    \Sigma = \Gamma^{-1}.
\end{equation}
The $a$\textsuperscript{th} diagonal entry of $\Sigma$ is a lower bound on the variance in the estimation of the parameter $\theta_a$,
\begin{equation}
\label{eq:cramer_rao}
    \langle \Delta\theta_a^2\rangle_n = \Sigma_{aa},
\end{equation}
where here the angled brackets and subscript $n$ indicates an averaging over detector noise realizations. This lower bound is referred to as the \textit{Cram\'er-Rao (CR) bound}, which is the minimum standard deviation that can be achieved with an unbiased estimator~\cite{Cutler_1994, Sathyaprakash_2009, Poisson_1995}. We will also make use of the correlation coefficient $\mathcal{C}^{ab}$, defined as
\begin{equation}
	\label{eq:correlation_coefficient}
	\mathcal{C}^{ab} = \Sigma^{ab}/\sqrt{\Sigma^{aa}\,\Sigma^{bb}}.
\end{equation}
This quantifies how errors in estimating different parameters are correlated \cite{Cutler_1994, Poisson_1995} and it is useful in identifying parameter degeneracy.

\subsection{Detector Sensitivities}
\label{sec:PSDs}

\begin{figure}[h]
    \centering
    \includegraphics[width=\columnwidth]{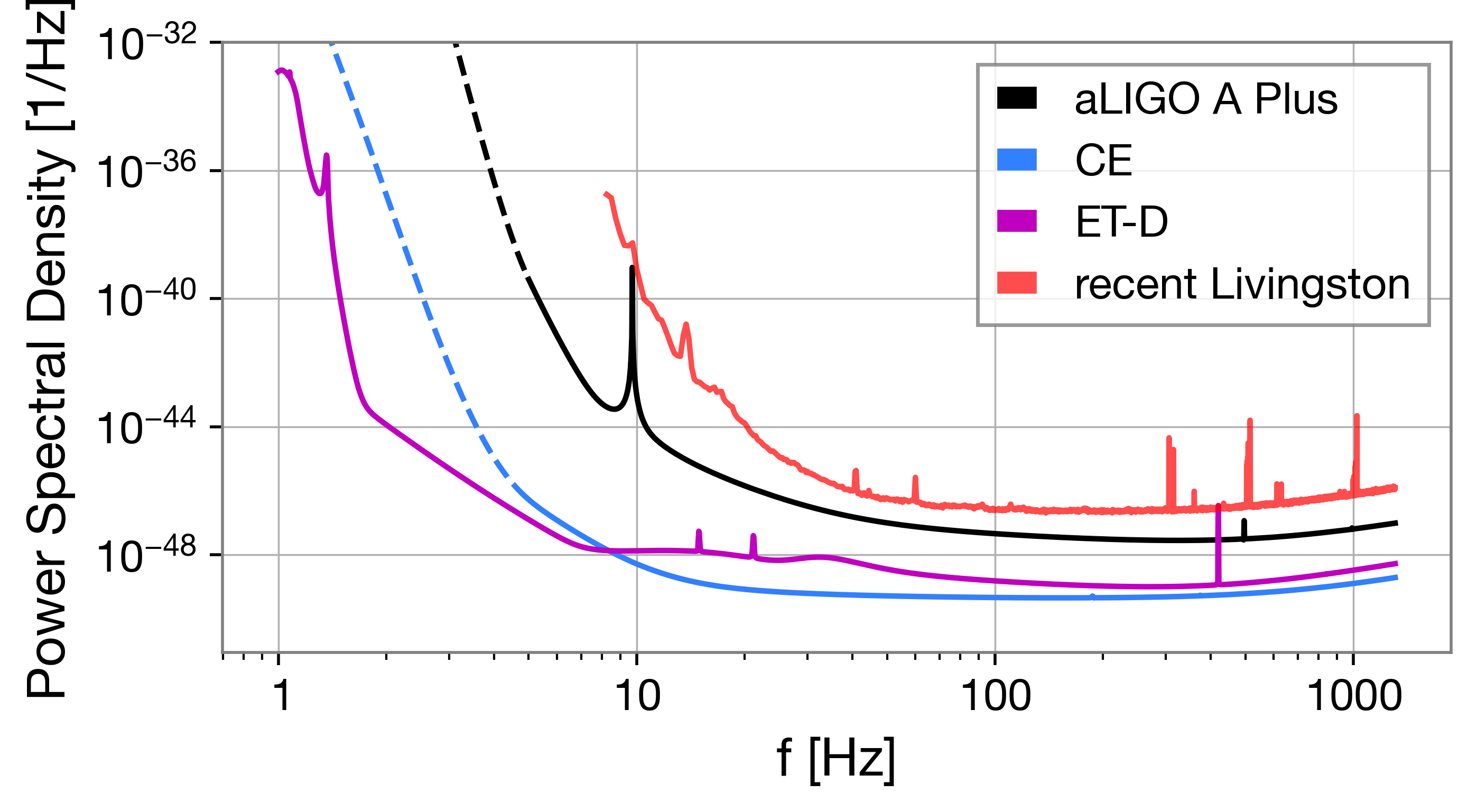}
    \caption{PSD projections used in our calculations for three third-generation ground-based GW detectors: aLIGO A+ (black), ET-D (magenta), and CE (blue), and an estimation for Livingston using recent data (red). References to the source of these PSD curves are given in the main text. A model for ET-D's PSD is available down to $1\,$Hz, while the curves for aLIGO A+ and CE are provided down to $5\text{ Hz}$; we have been extrapolated them to $1\text{ Hz}$ (dashed curves).  See text for how the recent Livingston PSD is obtained.}
    \label{fig:PSDs}
\end{figure}

The low frequency portion of the GW signal has the most information about CoM acceleration (see Eq. \eqref{eq:model} and Table \ref{tab:num_cycles}), which implies that third-generation detectors, designed to have reduced low-frequency noise (see Fig. \ref{fig:PSDs}), are particularly promising sites to detect $\alpha$. We therefore consider advanced LIGO A Plus (aLIGO A+) \cite{aLIGO_APlus_PSD}, Cosmic Explorer (CE) \cite{CE_PSD}, and Einstein Telescope (ET) \cite{ET_PSD}.

For ET, we use the ET-D sensitivity curve (see \cite{Hild_2011} for a discussion of the various ET configurations). The available PSD curve for ET-D goes down to a frequency of $1\,$Hz. However, sensitivity projections for both CE and aLIGO A+ are only available down to $5\,$Hz. We numerically extrapolate the logarithm of the aLIGO A+ and CE PSD curves down to $1\text{ Hz}$.  We use a quadratic extrapolation for aLIGO A+ and a cubic extrapolation for CE. 
We nevertheless find that using these extrapolated PSDs below 5 Hz make insignificant differences in the numerical results of the Fisher forecast because the noise level rapidly increases by several orders of magnitude over a few Hz.

In Figure \ref{fig:PSDs}, these PSDs are shown with aLIGO A+ in black, ET-D in purple, CE in blue, and an estimate for the recent LIGO Livingston PSD in red. The solid curves are official projections and the dashed curves are our extrapolations to very low frequencies.
The LIGO Livingston PSD is estimated with \texttt{gwpy.timeseries}~\cite{gwpy} and built-in routines for PSD estimation on a strain data series in the Livingston detector (we use 2048 seconds of data beginning 1 second after GW200316).  In the next section, we use these PSDs to calculate the covariance matrix Eq.~\eqref{eq:covariance} and thus forecast the detectability of the acceleration parameter $\alpha$.

\section{Results and Discussion}
\label{sec:detectability_analysis}

In this Section, we provide the results of the Fisher forecast. We begin with Section \ref{sec:rough_scaling}, in which we provide an analytic explanation about how $\Delta\alpha$ is expected to scale as a function of other parameters.  We continue in Section \ref{sec:numerical_results} by presenting the fully numerical Fisher results, including the CR bound for $\alpha$ and discussing the degeneracy between $\alpha$ and other parameters. We will fit the numerical results for the CR bound of $\alpha$ to simple analytical formulae as motivated by the results of Sec.~\ref{sec:rough_scaling}.

\subsection{Analytic Scaling}
\label{sec:rough_scaling}

Here, we provide an argument for how $\Delta\alpha$ is expected to scale with other parameters.  According to Eqs.~\eqref{eq:model}, \eqref{eq:Fisher}, and \eqref{eq:coefficient}, we see that
\begin{equation}
	\label{eq:rough_scaling}
	\frac{\Delta \alpha}{\alpha}\propto \frac{\mathcal{M}^{10/3}}{\rho\,\alpha},
\end{equation}
where $\rho$ is the SNR. This is derived by considering $\Delta\alpha/\alpha = \sqrt{\Sigma_{\ln\alpha\ln\alpha}} \approx 1/\sqrt{\Gamma_{\ln\alpha\ln\alpha}}$. Considering just the leading order term, we have $\Gamma_{\ln\alpha\ln\alpha}\propto \alpha^2\mathcal{A}^2\,A_0^2\,x^{-13}\propto \alpha^2\rho^2/\mathcal{M}^{20/3}$ (recall that $\mathcal{A}$ is the frequency domain signal amplitude, Eq.~\eqref{eq:total_waveform}) \cite{Cutler_1993}. This relation is inexact for a few reasons. First, inverting the Fisher matrix $\Gamma$ to obtain $\Sigma$ depends on the off-diagonal elements of $\Gamma$, as well as the higher-order corrections in the phase that are proportional to the $A_{n}$ coefficients. Second, evaluating the Fisher matrix requires integrating between a minimum frequency and the ISCO frequency (see the paragraph below Eq.~\eqref{eq:Fisher}), and the ISCO frequency itself is a function of the chirp mass $\mathcal{M}$. Nonetheless, we find that the analytic scaling result is a fair approximation (see Fig. \ref{fig:results_all_params}).

This relation provides a few noteworthy insights.  First, it implies that $\Delta\alpha$ is roughly constant at fixed SNR as a function of $\alpha$. This is consistent with Eq. \eqref{eq:model} because we consider only $O\left(\alpha\right)$ corrections in our model; differentiating $h(f)$ in Eq.~\eqref{eq:total_waveform} with respect to $\alpha$ does not bring any additional powers of $\alpha$ into the integrand of Eq.~\eqref{eq:Fisher}. Second, it implies that the CoM acceleration is most detectable for a binary with a small chirp mass.

\subsection{Numerical Results}
\label{sec:numerical_results}

We now report our numerical results for the Fisher matrix calculation on the entire set of parameters, $\left\{\ln\alpha,\,\ln\mathcal{M},\,\ln\mu,\,\chi_1,\,\chi_2,\,t_c,\,\phi_c\right\}$. We also provide results that correspond to some of these parameters held fixed.  We use the logarithms of $\alpha$, $\mathcal{M}$ and $\mu$ to avoid ill-conditioned matrices (see \cite{Iacovelli_2022} for a discussion of when ill-conditioning might occur when numerically evaluating the Fisher matrix). We set a fiducial SNR $\rho=10$, which corresponds to the majority of detectable GW events close to the detection threshold of matched filtering. All parameter derivatives are evaluated for equal mass binaries with $\mathcal{M}_{\rm bNs}/2 \leq \hat{\mathcal{M}} \leq 30\,M_{\odot}$ (where $\mathcal{M}_{\rm bNs}$ is the chirp mass for an neutron star binary with $m_1 = m_2 = 1.4 M_{\odot}$), $\hat{\mu} = 2^{-4/5} \hat{\mathcal{M}}$ corresponding to an equal mass binary, and $\hat{\chi}_1 = \hat{\chi}_2 = \hat{t}_c = \hat{\phi}_c = 0$ (we hereafter drop the hats). Note that $\mathcal{M}_{\text{bNs}}/2$ corresponds to a binary with component masses $m\approx 0.7\,M_{\odot}$. In agreement with Eq.~\eqref{eq:rough_scaling}, we find that varying the true value of $\alpha$ hardly affects the CR bound. Even changing $\alpha$ by orders of magnitude only alters the value of $\Delta\alpha$ by, at most, $0.1\%$, except when $\alpha$ have very large values $\sim 10^{-4}\text{ s}^{-1}$ and our perturbative phase model becomes inapplicable.

Although inexact, Eq.~\eqref{eq:rough_scaling} provides useful analytic guidance for $\Delta \alpha$ as a function of $\mathcal{M}$. We therefore fit
\begin{equation}
	\label{eq:fit}
	\Delta\alpha_i = c_i \left(\frac{10}{\rho}\right)\mathcal{M}^{e_i},
\end{equation}
to our numerical Fisher forecast for $\Delta\alpha = \sqrt{\Sigma_{\alpha \alpha}}$ according to Eq. \eqref{eq:covariance}. We expect that the exponents $e_i$ will equal $10/3$ only when all other parameters are fixed.  This is because the CR bound is calculated by inverting the entire Fisher matrix, whereas Eq. \eqref{eq:rough_scaling} comes from fixing all other parameters, considering the Fisher matrix to be a scalar, $\Gamma_{\alpha\alpha}$.  The parameters to be fit, $c_i$ and $e_i$, depend on the specific detector under consideration, which we label using the index $i$. Note that the units of $c_i$ is $\text{s}^{-1-e_i}$ and $e_i$ is dimensionless.

For the sensitivities given in Fig.~\ref{fig:PSDs}, the obtained values for $c_i$ and $e_i$ are shown in Table \ref{tab:fits} for the full range $\mathcal{M}_{\text{bNS}}/2\leq \mathcal{M} \leq 30\,M_{\odot}$; each tabular block in Table \ref{tab:fits} corresponds to different parameters being fixed, which is implemented by removing from the Fisher matrix rows and columns corresponding to those parameters.  As mentioned above, the exponents $e_i$ approach $10/3$ when other parameters are fixed, although they deviate from this value by, at most, $10\%$. We see that $c_i$ decreases by a factor of two depending on whether or not aligned spin components are included as parameters in the waveform model.


\begin{table}
	\centering
\begin{tabular}{|c|c|c|}
	\hline
	det. & $c_i$ [$\text{s}^{-1-e_i}$] & $e_i$\\
	\hline\hline
	\multicolumn{3}{|c|}{no fixed parameters}\\
	\hline
	aLIGO A+ & $4.31\times 10^{-8}$ & $3.60$  \\
	ET & $1.26\times 10^{-10}$ & $3.45$ \\
	CE & $2.82\times 10^{-9}$ & $3.53$ \\
	\hline\hline
	\multicolumn{3}{|c|}{fix $\chi_1=\chi_2=0$}\\
	\hline
	aLIGO A+ & $2.32\times 10^{-8}$ & $3.38$  \\
	ET & $6.95\times 10^{-11}$ & $3.36$ \\
	CE & $1.32\times 10^{-9}$ & $3.41$ \\
	\hline\hline
	\multicolumn{3}{|c|}{fix $\chi_1=\chi_2=t_c=\phi_c=0$}\\
	\hline
	aLIGO A+ & $2.03\times 10^{-8}$ & $3.27$ \\
	ET & $6.31\times 10^{-11}$ & $3.30$ \\
	CE & $1.10\times 10^{-9}$ & $3.31$ \\
	\hline
\end{tabular}
	\caption{Parameter values from fitting the empirical formula Eq. \eqref{eq:fit} to the numerical CR bound for $\Delta \alpha$, for an event with a fiducial SNR $\rho=10$ and $\alpha=0$.  We have used logarithmically spaced values $\mathcal{M}$ (see Fig. \ref{fig:results_all_params}) and fit the natural logarithm of Eq. \eqref{eq:fit} to $\ln\Delta\alpha$ because the CR bound varies by orders of magnitude as a function of $\mathcal{M}$. In the top block, all parameters are given uniform priors and are marginalized over the parameter space; in the middle block, $\chi_1$ and $\chi_2$ are fixed to zero; in the bottom block, $\chi_1, \chi_2, t_c$, and $\phi_c$ are all fixed to be zero.\label{tab:fits}}
\end{table}

\begin{figure}
	\centering
	\includegraphics[width=\columnwidth]{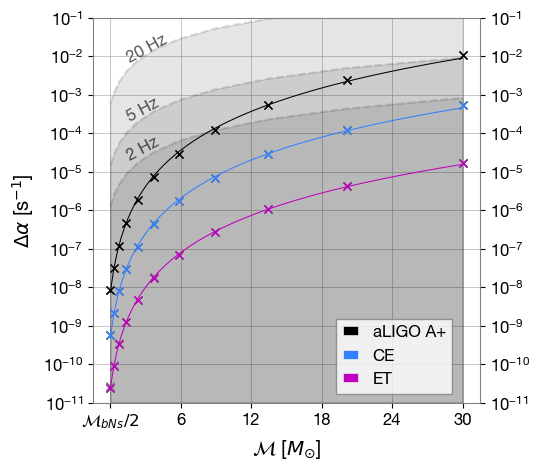}
	\caption{CR bound for $\Delta\alpha$ for a fiducial SNR $\rho=10$, evaluated as a function of the chirp mass $\mathcal{M}$ and with $f_{\rm min} = 1\text{ Hz}$; the regions shaded in gray denote regions in which terms $\mathcal{O}\left(\alpha^2\right)$ can be safely neglected if the effective minimum observable frequency is $20\,$Hz, $5\,$Hz and $2\,$Hz, appropriate for aLIGO A+, CE and ET, respectively (see Appendix \ref{sec:linear_alpha_justification} for details on how the boundary of the shaded region is calculated based on the effective minimum observable frequency). The markers are the CR bound for each detector and under the null hypothesis $\alpha = 0$, as per Eq. \eqref{eq:cramer_rao}. The colored curves are numerical fits to the empirical fitting formula Eq. \eqref{eq:fit}. Behind the colored markers are gray markers (most noticeable for ET with low chirp mass), which are the CR bounds for $\alpha$ in the range $\left[10^{-11}, 10^{-6}\right]\text{ s}^{-1}$. Changing $\alpha$ over orders of magnitude alters the CR bound by at most $0.1\%$, but usually much less. As a result, the gray markers are hardly visible in this plot. All results here correspond to a Fisher analysis centered at parameter values $\mu = 2^{-4/5}\mathcal{M}$ and $\chi_1 = \chi_2 = t_c = \phi_c = 0$.}
	\label{fig:results_all_params}
\end{figure}

The numerical results are shown, alongside the corresponding fitting formula of Eq.~\eqref{eq:fit}, in Figure \ref{fig:results_all_params}. The colored markers show the CR bounds for $\Delta\alpha$, from Eq. \eqref{eq:cramer_rao}, as a function of the chirp mass $\mathcal{M}$ and evaluated for $\alpha=0$. Results are shown for aLIGO A+, CE, and ET with fits to the empirical formula, Eq. \eqref{eq:fit}, shown with the colored lines. The fits show excellent agreement with the CR bounds. The gray markers behind the colored markers (see, for instance, the ET data at low chirp masses) are the CR bounds for nonzero acceleration values $\alpha = 10^{-11},\,\dots\,, \, 10^{-6}\,{\rm s}^{-1}$. These heavily overlap with the colored markers because the CR bounds change by, at most $0.1\%$, as a function of $\alpha$. The gray shades enclose the parameter space where quadratic corrections $O\left(\alpha^2\right)$ to the GW phase in Eq. \eqref{eq:model} are smaller than the linear contribution for effective minimum observable frequencies at $20\,$Hz, $5\,$Hz and $2\,$Hz. These values are appropriate for detection at aLIGO A+, CE, and ET, respectively. See Appendix \ref{sec:linear_alpha_justification} for details on how this frequency affects the validity of neglecting terms quadratic in $\alpha$. For all three detectors and masses shown, the detection limit calculated using our linear model is within the validity range. Note that the Fisher analysis is carried out for $f_{\rm min} = 1\text{ Hz}$ for all detectors.

A detection of CoM acceleration is possible when $\Delta\alpha/\alpha < 1$ for nonzero $\alpha$; otherwise, a clear detection is not possible. We see that for light chirp masses $\mathcal{M}_{\text{bNs}}/2\,\leq\mathcal{M}\leq 2\,M_\odot$ and with a moderate SNR $\rho=10$, detections can be made at aLIGO A+ for CoM acceleration $\alpha \sim 10^{-8}\text{ s}^{-1}$, whereas for stellar mass binaries, a CoM acceleration $\alpha\gtrsim 10^{-6}\,{\rm s}^{-1}$ is detectable.

Owing to excellent sensitivities at low frequencies, both CE and ET hold promise in detecting CoM acceleration.  The CR bound for $\alpha$ strongly depends on the chirp mass, as suggested by Eq. \eqref{eq:rough_scaling}.  For lighter sources with chirp masses less than $3\, M_{\odot}$, ET will be sensitive to $\alpha\sim (10^{-11}$--$10^{-9})\,\text{ s}^{-1}$, and CE will be sensitive to $\alpha\sim (10^{-9}$--$10^{-7})\,\text{s}^{-1}$.  For heavier BBH sources, ET will be sensitive to $\alpha\sim (10^{-8}$--$10^{-6})\,\text{ s}^{-1}$ and CE will be sensitive to $\alpha\sim (10^{-7}$--$10^{-5})\,\text{s}^{-1}$.   For GW events with higher SNR values, lower values of $\alpha$ can be detected according to Eq. \eqref{eq:fit}.

Incorporating Eq. \eqref{eq:alpha_value}, we can use the CR bounds of Fig. \ref{fig:results_all_params} to determine values of the tertiary mass and distance for which CoM acceleration is detectable.  ET is sensitive to a tertiary perturber with mass and distance satisfying $\left(M_3/M_\odot\right)\left(1 \text{ AU}/r\right)^2\gtrapprox 10/\text{SNR}$, CE can detect if this ratio is $\gtrapprox 10^{3}/\text{SNR}$, and aLIGO A+ for $\gtrapprox 10^4/\text{SNR}$. Therefore, our analysis indicates that detecting CoM acceleration with the third-generation detectors on the ground is possible for a stellar-mass tertiary in compact bound systems with a size $\lesssim 1\,$AU.

The above results can be gleaned from Table \ref{tab:fits}. That table additionally shows how the detectability of $\alpha$ improves if the spins of the compact objects are fixed (for instance if we believe the binary compact objects are non-spinning). In this case, the CR bound is improved by about a factor of $2$, and the possibility of detecting the effect of a CoM acceleration becomes even more promising.

The degeneracy between $\alpha$ and the other parameters can be quantified through the correlation coefficients, which are defined in Eq.~\eqref{eq:correlation_coefficient}. These dimensionless numbers, $\mathcal{C}_{ab}$, are between $-1$ and $1$ by definition and indicate the extent to which the errors in estimating two different parameters $\theta^a$ and $\theta^b$ are correlated. Typically, values $|\mathcal{C}_{ab}| > 0.9$ are taken to be indications of strong correlations and imply that a linear combination of the two parameters can be measured significantly more accurately than what can be done for $\theta^a$ and $\theta^b$ separately \cite{Cutler_1994}.  Nonetheless, the correlations appearing in Table \ref{tab:correlation_coefficients} imply that excluding $\alpha$ from data analysis of an event whose true $\alpha$ is nonzero can bias the inferred parameters (the extent to which parameters are biased is discussed in Appendix ~\ref{sec:injection}).

\begin{table}
    \centering
\begin{tabular}{|c|c|c|c|c|c|c|} 
\hline
\backslashbox{det.}{param.} & $\mathcal{M}$ & $\mu$ & $\chi_1$ & $\chi_2$ & $t_c$ & $\phi_c$\\\hline
aLIGO A+ & 0.86 & -0.78 & -0.75 & 0.75 & 0.62 & -0.72 \\ \hline 
CE & 0.88 & -0.81 & -0.74 & 0.74 & 0.58 & -0.73 \\ \hline 
ET & 0.84 & -0.76 & -0.66 & 0.66 & 0.50 & -0.66 \\ \hline 
\end{tabular}
    \caption{Correlation coefficients, $\mathcal{C}_{\alpha \theta^b}$ (defined in Eq.~\eqref{eq:correlation_coefficient}), between the CoM acceleration parameter $\alpha$ and parameter $\theta^b$, evaluated for each of the three third-generation GW detectors, for $\mathcal{M} = \mathcal{M}_{\text{bNS}}$, and $\alpha=0$. We see that $\alpha$ has large correlation coefficients with the other parameters, with $\mathcal{M}$ having the largest correlation of $C_{\alpha\mathcal{M}}\approx 0.8$ across all detectors. These correlation coefficients are sufficiently large to bias inferred parameter values away from their true values if a phase model without accounting for the effect of CoM acceleration is used applied to an acceleration GW source, see text.}
    \label{tab:correlation_coefficients}
\end{table}

Values of $\mathcal{C}_{\alpha \theta^b}$, where the second index runs over all parameters other than $\alpha$, are shown for each of the three third-generation ground-based detectors in Figure \ref{tab:correlation_coefficients}, evaluated for $\alpha=0$ (changing the value of $\alpha$ by orders of magnitude hardly affects the correlation coefficients). It is clear that all parameters show significant degeneracy with $\alpha$ to similar degrees, with $0.5\leq |\mathcal{C}_{\alpha \theta^b}|\leq 0.83$. 
It is worth noting that such degeneracy is significantly milder than that between the chirp mass $\mathcal{M}$, the reduced mass $\mu$, and component spins, which tends to be equal to $\pm 1$ up to three or four significant digits (e.g. \cite{Poisson_1995}). The mild degeneracy between $\alpha$ and the other parameters comes from the fact that $\alpha$ corrects the waveform phase at $\mathcal{O}\left(f^{-8/3}\right)$ relative to the usual PN terms, which depend on the various other parameters.  In other words, the $\alpha$ correction to the waveform phase scales uniquely with frequency compared to the PN terms, which provide information on all other parameters.  This weak degeneracy implies that if one observes an event where the source has a CoM acceleration, but a non-acceleration waveform model is used for parameter inference, the inferred component masses and spins will deviate from their true values.  This statement is explored through both an injection analysis and a semi-analytic argument in Appendix \ref{sec:injection}.

\section{Conclusion}
\label{sec:concl}

We have presented an analytic frequency-domain model for the gravitational waveform phase of an inspiraling compact binary system whose CoM is accelerating relative to the observer. Our model is valid to first order in the acceleration and it can be easily generalized to incorporate higher order effects. See Appendix \ref{sec:derivation} for a derivation of our model, including how to incorporate higher order effects, and Appendix \ref{sec:linear_alpha_justification} for a discussion of the range of validity of the linear approximation. We have shown that the acceleration appears as a correction at $\mathcal{O}\left(f^{-8/3}\right)$ compared to the intrinsic PN phase terms, see Eq. \eqref{eq:model}. Therefore, improvements of detector sensitivity at low frequencies at ground-based GW observatories will be pivotal for the detection of CoM acceleration. We have demonstrated, in particular, that the Einstein Telescope's focus on mitigating low-frequency seismic noise with the underground design, see Fig. \ref{fig:PSDs}, will make it the ideal detector for CoM acceleration, see Fig. \ref{fig:results_all_params} and Table \ref{tab:fits}. This paper presents the first analytic phase model up to 3PN order, which can readily be implemented in existing parameter estimation suites.

Through a Fisher information calculation, we have constrained the values of SNR and CoM acceleration that are detectable for third-generation GW detectors on the ground.  The main conclusions of our paper can be summarized with Eq. \eqref{eq:alpha_value}, Eq. \eqref{eq:fit}, Table \ref{tab:fits}, and Fig. \ref{fig:results_all_params}.  These imply that for binaries with chirp mass $\mathcal{M}_{\text{bNs}}/2$, ET can detect tertiary perturbers with mass and distance satisfying $\left(M_3/M_\odot\right)\left(1 \text{ AU}/r\right)^2\gtrapprox 10/\text{SNR}$, CE can detect if this ratio is $\gtrapprox 10^{3}/\text{SNR}$, and aLIGO A+ for $\gtrapprox 10^4/\text{SNR}$.  

The imprint of CoM acceleration in the GW signal will be one of the promising new signatures to look for at third-generation ground-based GW detectors \cite{Punturo_2010}. We have created a GW template incorporating this effect to help usher in the detection of hierarchical triple compact object systems with GWs.  As such, our model is a simple case in which two compact objects inspiral as they, together, orbit a massive tertiary body. This model can be further built upon by including Newtonian and post-Newtonian three-body perturbations, as well as eccentricity.  Eccentricity is of particular interest because hierarchical three-body systems are known to undergo Kozai-Lidov oscillations, in which an inner binary's eccentricity oscillates in coherently with the outer binary's inclination in such a way that the total angular momentum of the system is conserved~\cite{Naoz_2016}. In the context of our model, a nonzero CoM for an eccentric binary would imply a smoking gun for a triple system.  These will be directions of future research.

\section*{Acknowledgment}

We thank the anonymous referee whose suggestions expanded our analysis.  NL acknowledges support from NSF Physics Frontier Center Award 2020275. LD acknowledges research grant support from the Alfred P. Sloan Foundation (Award Number FG-2021-16495), and support of Frank and Karen Dabby STEM Fund in the Society of Hellman Fellows.

\appendix
\section{Outline of Derivation}
\label{sec:derivation}

Here we outline how to obtain the GW phase starting from PN expressions of the energy flux radiated from a circular binary, $\mathcal{F}$, and the bound state energy, $\mathcal{E}$:

\begin{equation*}
    \mathcal{F} = x^{5/2}\,\sum_i F_{i/2}\,x^{i/2},\,\,\,\, \mathcal{E} = \sum_i\,E_{i/2}\,x^{i/2},
\end{equation*}

\noindent where $x = (M\,\omega)^{2/3}$ is the usual PN expansion parameter and $\omega$ is the angular orbital frequency.  See Eqs. (232) and (314) of \cite{Blanchet_2014} for values of $F_{i/2}$ and $E_{i/2}$ for non-spinning sources and Eqs. (414) and (415) of \cite{Blanchet_2014} for spin-orbit coupling. We neglect all terms quadratic in the spins, which arise from the spin-spin dynamics at leading-order and spin-orbit dynamics at next-to-leading-order.

The orbital dynamics are then determined by radiation reaction considerations~\cite{Poisson_2014},
\begin{equation}
    \frac{\dd \mathcal{E}}{\dd t} = -\mathcal{F}.
\end{equation}
The chain rule gives $\dd \mathcal{E}/\dd t = (\dd \mathcal{E}/\dd x)/(\dd x/\dd t)$ and we can solve for $t(x)$ by integrating
\begin{equation}
\label{eq:t_of_x}
    t(x) - t_c = \int_{\infty}^{x}\frac{\dd  \mathcal{E}(x')/\dd x'}{\mathcal{F}(x')}\,\dd x'.
\end{equation}
This can be easily evaluated by expanding the integrand as PN expansion in $x$.  Note that this expression can be inverted as a power series in order to obtain $x(t)$.

The orbital phase is
\begin{equation}
\label{eq:phi_t}
    \phi(x) - \phi_c = \int_{t_c}^{t}\,\omega(t')\,\dd t' =  \frac{1}{M}\int_{t_c}^{t}\,x^{3/2}\,\dd t'.
\end{equation}
To evaluate this expression, we invert $t(x)$ given by Eq. \eqref{eq:t_of_x} to obtain

\begin{align}
    x(t) =& \frac{1}{4}\Theta^{-1/4} + \frac{743 + 924\,\nu}{16128}\,\Theta^{-1/2}\nonumber\\
    &+ \left(-\frac{\pi}{20} + \frac{1}{960\, M^2}\left\{\left[75\, M\,\Delta\, m_2 + 188\, m_2^2\right]\,\chi_2\right.\right.\nonumber\\
    &\left.\left.\hspace{50pt}+ \left[-75\, M\,\Delta\, m_2 + 188\, m_2^2\right]\,\chi_1\right\}\right)\Theta^{-5/8}\nonumber\\
    &\hspace{-30pt}+\frac{313328 + 512421\,\nu + 437472\,\nu^2}{16257024}\,\Theta^{-3/4}\nonumber\\
    &\hspace{-30pt}+\left(\frac{\pi\left(11891 + 3052\,\nu\right)}{215040}\right.\nonumber\\
    &\hspace{-10pt}\left.+ \frac{1}{2580480 M^2}\left\{\right.\right.\nonumber\\
    &\left.\left.\hspace{10pt}\left[5\,M\,\Delta\left(96473 - 6636\,\nu\right)\,m_2\right.\right.\right.\nonumber\\
    &\hspace{30pt}\left.\left.\left.+ 4\,\left(357923 - 5236\nu\right)\,m_2^2\right]\,\chi_2\right.\right.\nonumber\\
    &\left.\left.+ \left[5\, M\, \Delta\left(6636\, \nu-96473\right)\,m_1\right.\right.\right.\nonumber\\
    &\left.\left.\left.\hspace{30pt}+ 4\,\left(357923 - 5236\,\nu\right)m_1^2\right]\,\chi_1\right\}\vphantom{\frac{\pi}{215040}}\right)\Theta^{-7/8}\nonumber\\
    &\hspace{-20pt}+\left(\frac{1}{24034384281600}\left(\vphantom{\log\left(\frac{1}{4\Theta^{1/4}}\right)}-10052469856691\right.\right.\nonumber\\
    &\left.\left.+ 1530761379840\,\gamma_E + 1001432678400\, \pi^2 \right.\right.\nonumber\\
    &\left.\left.+ 24236159077900\, \nu - 
 882121363200\, \pi^2\, \nu\right.\right.\nonumber\\
 &\left.\left.- 206607970800\, \nu^2 + 
 462992376000\, \nu^3\right.\right.\nonumber\\
 &\left.\left.+ 3061522759680 \,\log 2 + \right.\right.\nonumber\\
 &\left.\left.
 765380689920 \log\left(\frac{1}{4\Theta^{1/4}}\right)\right)\right.\nonumber\\
 &\left.+ \frac{\pi}{364 M^2}\left\{\left[-45\, M\, \Delta\, m_2 - 118\, m_2^2\right]\,\chi_2\right.\right.\nonumber\\
 &\left.\left.\hspace{38pt}+ \left[45\, M\, \Delta\, m_1 - 118\, m_1^2\right]\,\chi_2\right\}\vphantom{\frac{1}{24034384281600}}\right)\Theta^{-1},
\end{align}
where $\Theta = \frac{\nu}{5 M}\left(t_c - t\right)$, $\gamma_E$ is the Euler-Mascheroni constant, $M = m_1 + m_2$, $\Delta = (m_1 - m_2)/M = \sqrt{1-4\nu}$, and $\nu = \mu/M$ is the symmetric mass ratio.  This expression can be compared to Eq. 316 in ~\cite{Blanchet_2014}, with which our result agrees; unlike Eq. 316 in ~\cite{Blanchet_2014}, we have included spin.

and then we can evaluate the above integral by expanding $x^{3/2}$ as a PN expansion in $t$.  Doing so yields $\phi(t)$; see Eq. (317) of \cite{Blanchet_2014} for the expression neglecting spin.

We obtain the waveform in the frequency domain by approximating the full Fourier transform,
\begin{equation}
\label{eq:fourier_transform}
    h(f) = \int_{-\infty}^{\infty} h(t)\,e^{2\pi i f t}\,\dd t,
\end{equation}
where $h(t) = A(t)\,e^{2\,i\,\phi(t)}$, via the stationary phase approximation (SPA)
\begin{equation}
\label{eq:stationary_phase_approximation}
    h(f)\approx \frac{1}{2}\,A\left(t^*(f)\right)\,\sqrt{\frac{\dd f}{\dd t}}\,\exp\left[2\pi\,f\,t^*(f) - \frac{\pi}{4} - 2\phi(t^*(f))\right],
\end{equation}
where $t^*(f)$ satisfies
\begin{equation}
\frac{\dd\phi(t)}{\dd t}\bigg\rvert_{t=t^*(f)} = \pi\,f,
\end{equation}
and it is understood that $\dd f/\dd t$ in Eq. \eqref{eq:stationary_phase_approximation} is evaluated at $t=t^*(f)$.

So far, we have not incorporated the CoM acceleration.  As described in the text, we consider the case in which CoM acceleration is completely decoupled from the binary's relative dynamics. We incorporate the effect by considering the time-delayed signal
\begin{align}
    \phi_{\alpha}(t) &= \phi_0\left(t + \alpha(t-T)^2\right),
\end{align}
and using this as the phase entering the Fourier transform of Eq. \eqref{eq:fourier_transform}. We now solve for $t_{\alpha}^*(f)$ that satisfies
\begin{equation}
    \label{eq:solve_for_tstar}
    \left[ 1+2\,\alpha\,(t^*(f)-T)\right]\,\frac{\dd\phi(t)}{\dd t}\bigg\rvert_{t=t^*(f)} = \pi\,f.
\end{equation}

Solving for $t^*(f)$ is then straightforward. We use the ansatz
\begin{align}
    t^*(f) = t_c - x^{-4}\,\big(t_0 + t_2\,x + t_3\,x^{3/2} &+ t_4\,x^2 + t_5\,x^{5/2}\nonumber\\
    \label{eq:tstar}
    &+ t_6\,x^3 + \alpha\,t_{A}(x) \big),
\end{align}
where $x=\left(\pi M f\right)^{2/3}$, the $t_{2n}$ are functions of the masses and spins, and $t_{A}(x)$ must be solved for via Eq. \eqref{eq:solve_for_tstar}.  At this point, letting $T = t_c$ greatly simplifies the resulting algebra.  With $t^*(f)$, we can now evaluate

\begin{multline}
\label{eq:SPA_phase}
    \psi_{\alpha}(f) = 2\pi\,f\,t^*(f) - \frac{\pi}{4} - 2\,\phi\left(t^*(f)\right)\\
    =2\pi\,f\,t_c - \phi_c - \frac{\pi}{4} + x^{-5/2}\sum_{n=0}^{6}x^{n/2}\left(P_{n} + S_n + \alpha\,A_n\,x^{-4}\right),
\end{multline}

and

\begin{align}
    P_0 &= \frac{3}{128\,\nu}\nonumber\\
    P_1 &= 0\nonumber\\
    P_2 &= \frac{5\left(743 + 924\,\nu\right)}{32256\,\nu}\nonumber\\
    P_3 &= -\frac{3\pi}{8\nu}\nonumber\\
    P_4 &= \frac{15293365 + 5040\,\nu\left(5429+4319\,\nu\right)}{21676032\,\nu}\nonumber\\
    P_5 &= -\frac{5 \pi}{64512 \nu} \left(-7729 + 1092\,\nu\right)\left(2+3\ln x\right)\nonumber\\
    P_6 &= \frac{1}{200286535680\,\nu}\left\{4527600 \left[(45633-102260\,\nu ) \nu ^2\right.\right.\nonumber\\
    &\hspace{20pt}\left.\left.+\,432 \pi ^2 (451\,\nu -512)\right]-24236159077900\,\nu \right.\nonumber\\
    &\hspace{20pt}\left.-\,1530761379840\,\gamma +11583231236531\right.\nonumber\\
    &\hspace{20pt}\left.-\,3061522759680 \ln 2 - 765380689920 \ln x\right\}\nonumber\\
    S_1 &= 0\nonumber\\
    S_2 &= 0\nonumber\\
    S_3 &= \frac{1}{128\, M^2 \,\nu}\left\{\left[188\,m_1^2 - 75M\Delta \,m_1\right]\chi_1\right.\nonumber\\
    &\hspace{20pt} \left.+\, \left[188\,m_2^2 + 75 M\Delta\, m_2\right]\chi_2\right\}\nonumber\\
    S_4 &= 0\nonumber\\
    S_5 &= \frac{5\left(2+3\ln x\right)}{193536\, M^2\, \nu}\left\{\left[\vphantom{m_1^2}27\,M\Delta \,m_2\left(-2783+56\nu\right)\right.\right.\nonumber\\
    &\hspace{75pt}\left.\left.-22\,m_2^2\left(10079 + 252\nu\right)\right]\chi_2\right.\nonumber\\
    &\hspace{66pt}\left. +\left[27\,M\Delta\,m_1\left(2783-56\nu\right)\right.\right.\nonumber\\
    &\hspace{75pt}\left.\left. -22\,m_1^2\left(10079 + 252\nu\right)\right]\chi_1\right\}\nonumber\\
    S_6 &= \frac{5\pi}{16\,M^2\, \nu}\left\{\left[118\,m_2^2 + 45 M \Delta\, m_2\right]\chi_2\right.\nonumber\\
    &\hspace{42pt}+\left.\left[118\,m_1^2 -45M \Delta\, m_1\right]\chi_1\right\}\nonumber\\
    A_0 &= \frac{25\,M}{32768\,\nu^2}\nonumber\\
    A_1 &= 0\nonumber\\
    A_2 &= \frac{25\,M\left(743 + 924\,\nu\right)}{4128768\,\nu^2}\nonumber\\
    A_3 &= -\frac{5M\,\pi}{512\,\nu^2}\nonumber\\
    &\hspace{20pt}+ \frac{5}{24576 \,M \,\nu^2}\left\{\left[-75 \,M\Delta\, m_1 + 188\,m_1^2\,\right]\chi_1 \right.\nonumber\\
    &\hspace{87pt}\left.+ \left[75 \,M\Delta\, m_2 + 188\, m_2^2\,\right]\chi_2\right\}\nonumber\\
    A_4 &= \frac{25\,M}{2774532096 \,\nu^2}\left(1755623\right.\nonumber\\
    &\left.\hspace{100pt}+ 112\, \nu\, (32633 + 23121 \,\nu)\right)\nonumber\\
    A_5 &= -\frac{5\,M\pi\left(20807 + 8036\,\nu\right)}{1376256\, \nu^2}\nonumber\\
    &\hspace{15pt}+ \frac{5}{4128768 \,M\, \nu^2}\left\{\left[-5M\Delta\, m_1\left(32477 + 8736\,\nu\right)\right.\right.\nonumber\\
    &\hspace{91pt}\left.\left.+ 14\, m_1^2\,\left(33049 + 8932\,\nu\right)\right]\chi_1\right.\nonumber\\
    &\hspace{80pt}\left.+\left[5\,M\Delta\, m_2\left(32477 + 8736\,\nu\right)\right.\right.\nonumber\\
    &\hspace{90pt}\left.\left.\left.+ 14 \,m_2^2\,\left(33049 + 8932\,\nu\right)\right]\chi_2\right.\right\}\nonumber\\
    \label{eq:coefficient}
    A_6 &= \frac{M}{46146017820672\,\nu^2}\left\{23100 \,\nu \,(3311653861 \right.\nonumber\\
    &\hspace{20pt} \left.+\, 84 \,\nu\, (2030687 + 1856036 \,\nu))\right.\nonumber\\
    &\hspace{20pt}\left.-\,234710784 \,\pi^2 \,(-18944 + 11275 \,\nu)\right.\nonumber\\
    &\hspace{20pt}\left.-\,28907482848623+ 4592284139520\, \gamma \right.\nonumber\\
    &\hspace{20pt}\left.+\,9184568279040 \ln2+ 2296142069760 \ln x \right\}\nonumber\\
    &+\frac{\pi}{6144\,M\,\nu^2}\left\{\left[1725 M\Delta\, m_1 - 4454 \,m_1^2\,\right]\chi_1\right.\nonumber\\
    &\hspace{54pt}\left.-\left[1725 M\Delta \,m_2 + 4454 \,m_2^2\,\right]\chi_2\right\}
\end{align}


\noindent Our expression agrees with the result of \cite{Tamanini_2020} (see their Eq. (A23), in which our $A_0$ appears).  Furthermore, our method can be compared to the numerical framework outlined by~\cite{Chamberlain_2019}, which considered a time-dependent time delay for an entire inspiral-merger-ringdown signal.

\section{Validity of Neglecting Quadratic Correction}
\label{sec:linear_alpha_justification}

We now address the values of $\alpha$ for which it is valid to truncate the expression
\begin{equation}
\label{eq:quadratic_correction}
    \psi_{\alpha}(f) = \psi_{\alpha}(f)\bigg\rvert_{\alpha=0} + \frac{\partial\psi}{\partial\alpha}\bigg\rvert_{\alpha=0}\alpha+\frac{1}{2}\frac{\partial^2\psi}{\partial\alpha^2}\bigg\rvert_{\alpha=0}\alpha^2 + O(\alpha^3)
\end{equation}

\noindent where the $\psi$ appearing here is the phase of the SPA waveform.  To make notation less cumbersome, we define $\psi_n = \partial^n\psi/\partial\alpha^n$.

We now quantify the range of $\alpha$ for which it is valid to neglect $\psi_2$, as we have done in Eq. \eqref{eq:model}. To derive $\psi_n$, we repeat the steps of Appendix \ref{sec:derivation} but we add a term $\alpha^2 t_B(x)$ in Eq. \eqref{eq:tstar}.  Doing so gives the quadratic correction

\begin{multline}
    \psi_2 = \alpha^2(B_0 x^{-21/2} + B_2 x^{-19/2} + B_3 x^{-9}\\ + B_4 x^{-17/2} + B_5 x^{-8} + B_6 x^{-15/2}),
\end{multline}

\noindent and

\begin{equation}
\begin{aligned}
    B_0 &= \frac{625\, M^2}{12582912\, \nu^3}\\
    B_2 &= \frac{1625 \,M^2\,(924 \,\nu +743)}{3170893824 \,\nu ^3}\\
    B_3 &= -\frac{25 \pi  M^2}{32768\, \nu^3}\\
    B_4 &= \frac{1375 \,M^2}{6392521949184\,\nu^3}\left\{7475065\right. \\
    &\left.\hspace{50pt} +2352 \,\nu \left(6997 + 4755\, \nu\right)\right\}\\
    B_5 &= -\frac{125\pi\, M^2}{1585446912\, \nu ^3}\left(86197 + 53676\,\nu\right)\\
    B_6 &= \frac{5\,M^2}{11813380562092032\,\nu^3}\left\{-82698671170559\right.\\
    &\hspace{20pt}\left.+ 13776852418560\, \gamma - 704132352 \,\pi^2 \left(-25088\right.\right.\\
    &\hspace{20pt}\left.+ 11275\, \nu\right)+ 69300 \,\nu \,(3572915219 + 84\, \nu \,\left(5819253\right.\\
    &\hspace{20pt}\left.+ 3592540\,\nu\right))\left.+ 27553704837120 \ln 2\right.\\
    &\hspace{20pt}\left.+ 6888426209280 \ln x\right\}.
\end{aligned}
\end{equation}

\noindent These expressions neglect spin contributions because our analysis has considered true binary values of $\chi_1 = \chi_2 = 0$.

Evidently, the quadratic terms in $\alpha$ are of $\mathcal{O}\left(f^{-8/3}\right)$ relative to the terms linear in $\alpha$ and are thus $\mathcal{O}\left(f^{-16/3}\right)$ relative to the non-acceleration terms.  They are, therefore, potentially impactful at low frequencies and while they may be important for large accelerations, our study has focused on the linear correction.

We have that the relation $\alpha\ll(\partial\phi/\partial\alpha)/(\partial^2\phi/\partial\alpha^2)$ can be written in the form

\begin{multline}
\label{eq:quadratic_to_linear}
   \alpha\ll\frac{\partial\phi/\partial\alpha}{\partial^2\phi/\partial\alpha^2} = x^4(C_0 + C_2 x + C_3 x^{3/2} + C_4 x^2\\ + C_5 x^{5/2} + C_6 x^3),
\end{multline}

\noindent as a PN expansion, with $C_{2n}$ being functions of $A_{n}$ and $B_{2n}$,

\begin{align}
	C_0 &= A_0/B_0\nonumber\\
	C_2 &= \left(A_2 B_0 - A_0 B_2\right)/B_0^2\nonumber\\
	C_3 &= \left(A_3 B_0 - A_0 B_3\right)/B_0^2\nonumber\\
 	C_4 &= \left(A_4 B_0^2 - A_2 B_2 B_0 + A_0 B_2^2 - A_0 B_0     B_4\right)/B_0^3\nonumber\\
	C_5 &= \left(A_5 B_0^2 - A_3 B_0 B_2 - A_2 B_0 B_3 + 2 A_0 B_2 B_3\right.\nonumber\\
        &\hspace{20pt}\left.- A_0 B_0 B_5\right)/B_0^3\nonumber\\
    C_6 &= \left(A_6 B_0^3 - A_4 B_0^2 B_2 + A_2 B_0 B_2^2 - A_0 B_2^3 - A_3 B_0^2 B_3\right.\nonumber\\
    &\hspace{20pt}\left.+ A_0 B_0 B_3^2 - A_2 B_0^2 B_4 + 2 A_0 B_0 B_2 B_4\right.\nonumber\\
    &\hspace{20pt}\left.- A_0 B_0^2 B_6\right)/B_0^4.
\end{align}

We can therefore get an upper bound for $\alpha$ by evaluating the RHS for the lightest binary considered here, with $m_1 = m_2 = 0.7 M_{\odot}$ (corresponding to a chirp mass $\mathcal{M}_{\text{bNs}}/2$), and a minimum observing frequency of $1 \text{ Hz}$.  Then we have that the RHS of Eq. \eqref{eq:quadratic_to_linear} is equal to $2.03\times 10^{-7}\text{ s}^{-1}$.  The largest term in Eq. \eqref{eq:quadratic_to_linear} is $x^4\, C_0$ and so, $\psi_1/\psi_2\propto \mathcal{M}^{5/3}$.  Therefore, the upper bound on $\alpha$ is closest to $0$ for small chirp mass and throughout our study, where the minimum $\mathcal{M}$ is $\mathcal{M}_{\text{bNS}}/2$, the smallest upper bound on $\alpha$ is thus $2.03\times 10^{-7}\text{ s}^{-1}$.  Note that the minimum observing frequency plays a role in this analysis, which we have fixed to $1\text{ Hz}$ in this Appendix.  In Fig. \ref{fig:results_all_params}, we present regions in which $\Delta\alpha\ll |\partial\phi/\partial\alpha|/|\partial^2\phi/\partial\alpha^2)|$ for a minimum observable frequency at $20\,$Hz, $5\,$Hz and $2\,$Hz, evaluated as a function of $\mathcal{M}$.

\begin{figure}[H]
    \centering
    \includegraphics[width=\columnwidth]{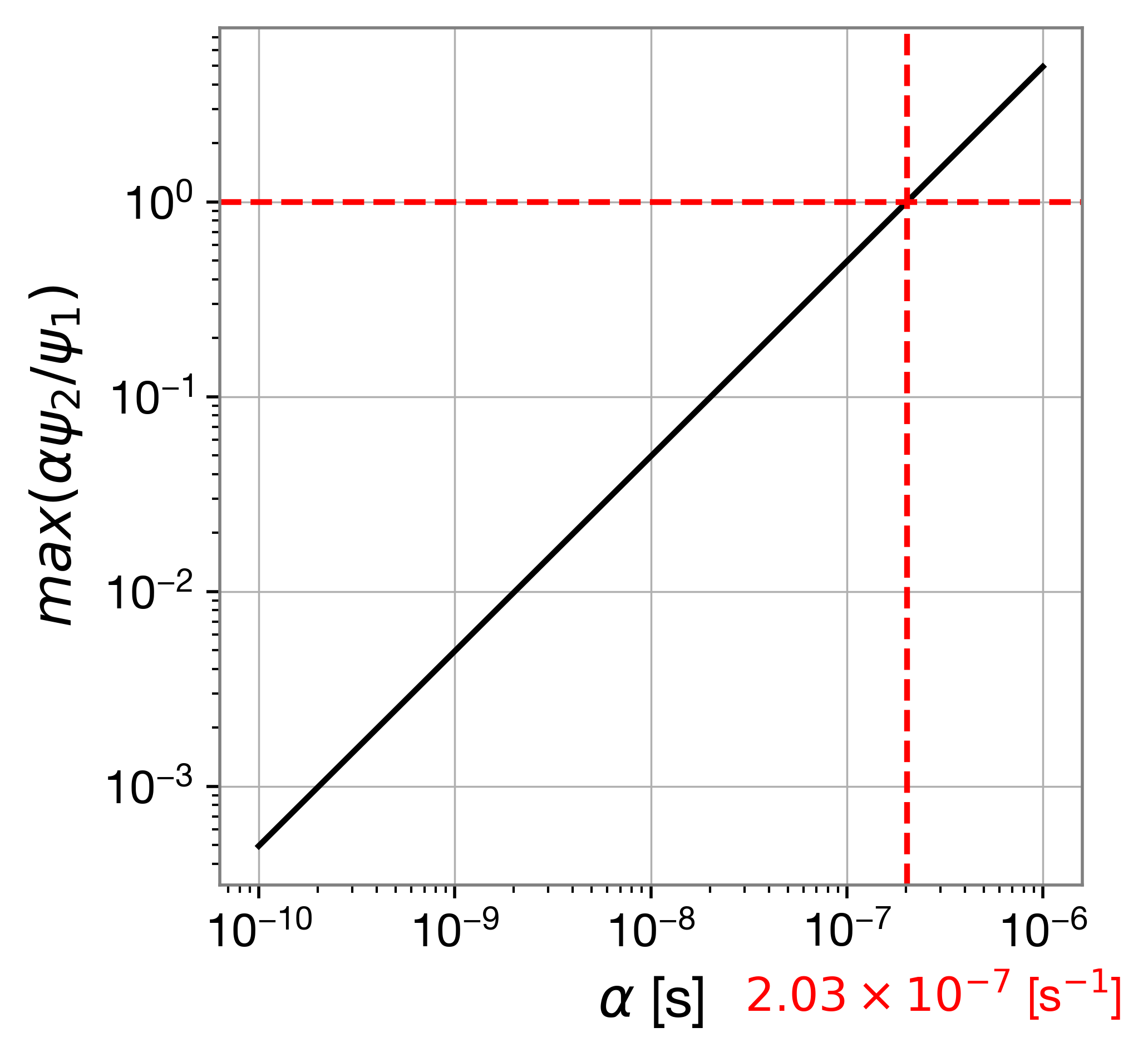}
    \caption{The maximum value of $\alpha\,\psi_2\,/\,\psi_1$, the ratio $\alpha^2\,\psi_{2}$ to $\alpha\,\psi_1$ as a function of $\alpha$ for fixed $\mathcal{M} = \mathcal{M}_{\text{bNS}}/2$.  By comparing the $O\left(\alpha^2\right)$ and $O\left(\alpha\right)$ corrections to our model, we see that the quadratic correction is much less than, and therefore negligible to, the linear correction as long as $\alpha \ll 10^{-7}\text{ s}^{-1}$.  Once $\alpha\sim O\left(10^{-7}\text{ s}^{-1}\right)$, then $\alpha^2\,\psi_2$ is within an order of magnitude of $\alpha\,\psi_1$ until they become equal at $\alpha\approx 2.03\times 10^{-7}\text{ s}^{-1}$.  Therefore, we set this value of $\alpha$ as the absolute maximum limit of $\alpha$ for which our model can be valid, this is marked in Fig. \ref{fig:results_all_params} by the gray regions.}
    \label{fig:quadratic_comparison}
\end{figure}

This result for $\mathcal{M} = \mathcal{M}_{\text{bNS}}/2$ is numerically confirmed in Figure \ref{fig:quadratic_comparison}, where we consider the maximum value of the ratio between the quadratic correction, $\alpha^2\,\psi_2$, and the linear correction, $\alpha\,\psi_1$ for fixed $\mathcal{M} = \mathcal{M}_{\text{bNS}}/2$.  We see that for small $\alpha$, until $\alpha\sim O\left(10^{-7}\text{ s}^{-1}\right)$, the quadratic correction is (at least) an order of magnitude smaller than the linear correction.  They become equal at $\alpha \approx 2.03\times 10^{-7}\text{ s}^{-1}$, which we therefore take as the maximum possible value for $\alpha$ at which our model is valid for the $\mathcal{M}_{\text{bNs}}/2$ system.  This is the value marked by the green region in Fig. \ref{fig:results_all_params} on the $\Delta\alpha$-axis, where it is given as a function of the chirp mass.

\section{Inferred Masses Deviate from True Values if Acceleration is Present but Unmodeled}
\label{sec:injection}

In this Appendix, we present an injection analysis that demonstrates how binary parameters can be biased if a non-accelerating model is used for parameter inference of an event with true $\alpha_{\rm tr}\neq 0$. In particular, we injected our template (cf. Eq. \eqref{eq:model}) with various values of the CoM acceleration, fixed component masses $m_1=m_2=1.4\,M_{\odot}$, and fixed aligned spin components $\chi_{1,\,z} = \chi_{2,\,z} = 0$, into mock noise for a detector with the projected ET-D PSD (cf. Fig. \ref{fig:PSDs}). We then apply the relative binning method for fast likelihood evaluation~\cite{Zackay_2018} and our model with $\alpha$ set to $0$ to infer the chirp mass, symmetric mass ratio, and component spins. We derive posterior probability distributions with the nested sampling Monte Carlo algorithm MLFriends (\cite{Buchner_2016, Buchner_2019}) using the UltraNest\footnote{\url{https://johannesbuchner.github.io/UltraNest/}} package \cite{Buchner_2021}.

\begin{figure*}
    \centering
    \includegraphics[width=\linewidth]{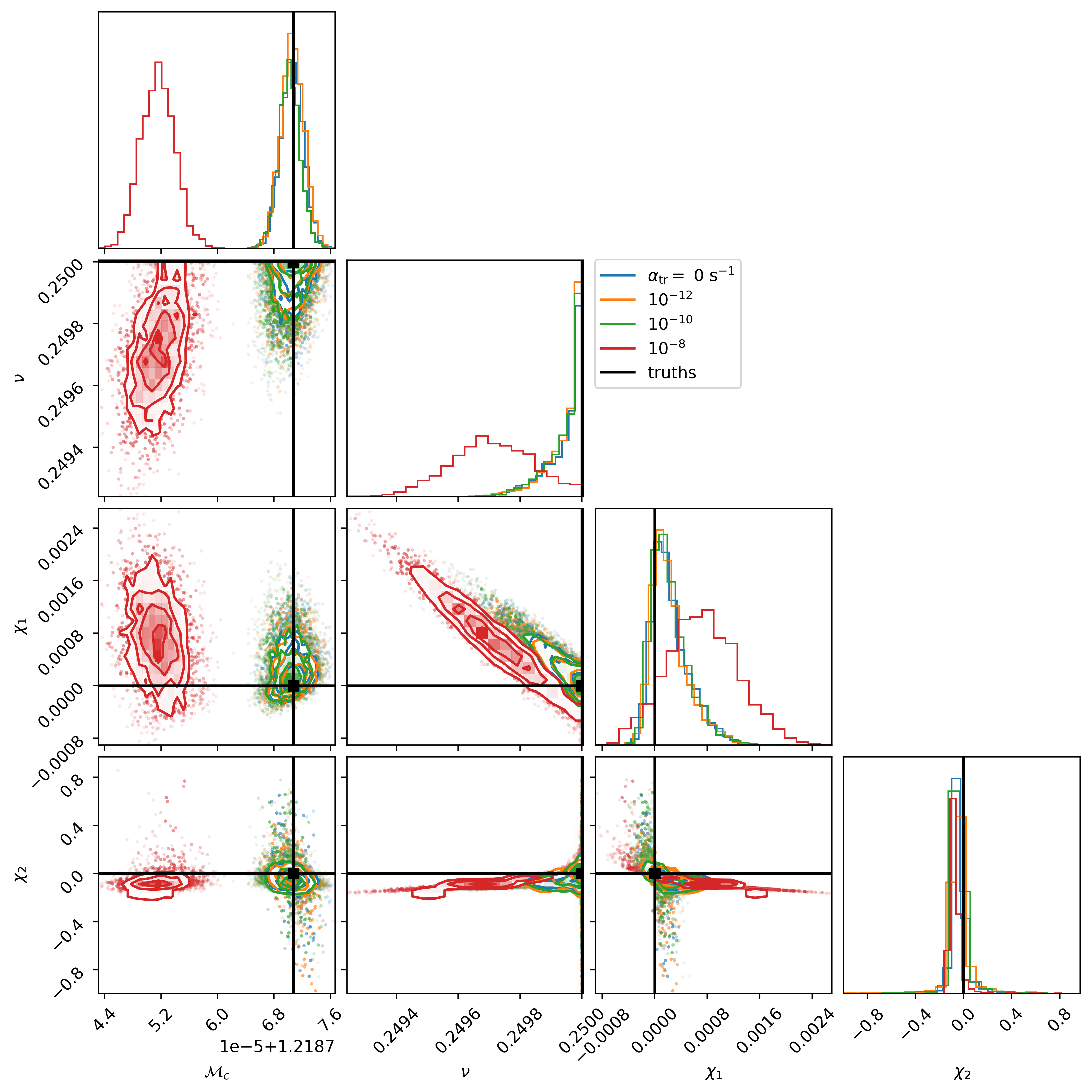}
    \caption{Posterior distributions of binary parameters when TaylorF2 (cf. Eq. \eqref{eq:model}) with $\alpha=0$ is used to analyze a signal with true $\alpha_{\rm tr}\neq 0$ given by the colors ($\alpha_{\rm tr}=0\text{ s}^{-1}$ (blue), $10^{-12}\text{ s}^{-1}$ (orange), $10^{-10}\text{ s}^{-1}$ (green), $10^{-8}\text{ s}^{-1}$ (red)) and the black solid lines denoting the parameters' true values. We see that for $\alpha\gtrsim 10^{-8}\text{ s}^{-1}$, the chirp mass can deviate by significantly more than 1$\sigma$, while $\nu$ deviates moderately from its true value. While the posterior distributions for the aligned spin components are centered at slightly nonzero values, they are still consistent with zero.}
    \label{fig:corners}
\end{figure*}

A corner plot~\cite{corner} of our analysis is shown in Figure \ref{fig:corners}. Curves of various colors correspond to different injected values of $\alpha_{\rm tr}$, $\alpha_{\rm tr}=0\text{ s}^{-1}$ (blue), $10^{-12}\text{ s}^{-1}$ (orange), $10^{-10}\text{ s}^{-1}$ (green), $10^{-8}\text{ s}^{-1}$ (red). The true values of the other parameters are marked by the solid black lines.  Each posterior distribution is derived by fixing $\alpha = 0\text{ s}^{-1}$ in our fitting model and using uniform priors for all parameters. We see that both the inferred chirp mass and symmetric mass ratio deviate by significantly more than 1$\sigma$ for CoM acceleration as large as $\alpha\gtrsim 10^{-8}\text{ s}^{-1}$.  For smaller values of $\alpha_{\rm tr}$, the inferred parameters are consistent with their true values. 

This brief analysis demonstrates that neglecting CoM acceleration can bias the inferred binary parameters for $\alpha_{\rm tr}$ as large as $10^{-8}\text{ s}^{-1}$. It is unlikely that neglecting CoM acceleration will qualitatively change the interpretation of the GW signal because the chirp mass is biased by about one part in a hundred thousand and the symmetric mas ratio is biased by a few parts in ten thousand.

\pagebreak

\bibliography{combib}
\bibliographystyle{apsrev4-1}

\end{document}